\documentclass[11pt, a4paper]{article}
\pdfoutput=1
\usepackage{jcappub}

\usepackage{amssymb,amsmath,latexsym,mathrsfs}
\usepackage{graphicx} 
\usepackage{bm}
\usepackage{url}
\usepackage{epsfig,graphicx,verbatim,xspace,multirow,mathtools}
\usepackage{array}
\usepackage{enumitem}

\usepackage[dvipsnames]{xcolor}

\newcommand{\zero}{{\color{white}{0}}}
 
\begin{document}


\title{Addendum: Refined bounds on MeV-scale thermal dark sectors from BBN and the CMB}

\author[a]{Nashwan Sabti,${}^1$\note{ORCID: \href{http://orcid.org/0000-0002-7924-546X}{0000-0002-7924-546X}}}
\emailAdd{nashwan.sabti@kcl.ac.uk}
\author[a]{James Alvey,${}^2$\note{ORCID: \href{https://orcid.org/0000-0003-2020-0803}{0000-0003-2020-0803}}}
\emailAdd{james.alvey@kcl.ac.uk}
\author[b]{Miguel Escudero,${}^3$\note{ORCID: \href{http://orcid.org/0000-0002-4487-8742}{0000-0002-4487-8742}}}
\emailAdd{miguel.escudero@tum.de}
\author[a]{Malcolm Fairbairn,${}^4$\note{ORCID: \href{https://orcid.org/0000-0002-0566-4127}{0000-0002-0566-4127}}}
\emailAdd{malcolm.fairbairn@kcl.ac.uk}
\author[c]{Diego Blas${}^5$\note{ORCID: \href{https://orcid.org/0000-0003-2646-0112}{0000-0003-2646-0112}}}
\emailAdd{dblas@ifae.es}

\affiliation[a]{Theoretical Particle Physics and Cosmology Group, Department of Physics,  \\
King's College London, Strand, London WC2R 2LS, UK}

\affiliation[b]{Physik-Department, Technische Universit{\"{a}}t, M{\"{u}}nchen, James-Franck-Stra{\ss}e, 85748 Garching, Germany}

\affiliation[c]{Grup  de  F\'isica  Te\`orica,  Departament  de  F\'isica, \\ Universitat  Aut\`onoma  de  Barcelona,  08193  Bellaterra, Spain\\
Institut de Fisica d’Altes Energies (IFAE),\\ The Barcelona Institute of Science and Technology, Campus UAB, 08193 Bellaterra, Spain}

\abstract{Very recently, the LUNA collaboration has reported a new measurement of the $d+p\to {}^{3}\text{He}+\gamma$ reaction rate, which plays an important role in the prediction of the primordial deuterium abundance at the time of BBN. This new measurement has triggered a new set of global BBN analyses within the context of the Standard Model. In this addendum to \href{https://doi.org/10.1088/1475-7516/2020/01/004}{JCAP 01 (2020) 004} (\href{https://arxiv.org/abs/1910.01649}{arXiv:1910.01649}), we consider the implications of these new results for our constraints on MeV-scale dark sectors. Importantly, we find that our bounds in the BBN-only and Planck-only analyses are insensitive to these updates. Similarly, we find that our constraints derived using BBN and CMB data simultaneously are not significantly modified for neutrinophilic particles. The bounds on electrophilic dark sector states, however, can vary moderately when combining BBN and CMB observations. We present updated results for all the relevant light dark sector states, calculated using the rates obtained by the leading groups performing standard BBN analyses.}
\maketitle

\noindent \textbf{Recent LUNA Results.} The LUNA collaboration has recently reported a very precise measurement of the $d+p\to {}^3\text{He} + \gamma$ cross-section at the energies relevant for Big Bang Nucleosynthesis (BBN) ($E \sim 30-300\,\mathrm{keV}$)~\cite{Mossa:2020gjc}. This quantity is particularly important for the prediction of the primordial deuterium abundance~\cite{Fields:2019pfx}. As a result, the three leading groups currently performing Standard BBN analyses have updated their predictions for the primordial deuterium abundance~\cite{Pitrou:2020etk,Pisanti:2020efz,Yeh:2020mgl}. These three groups agree on the form of the $d+p\to {}^3\text{He} + \gamma$ reaction rate as a function of temperature, and its contribution to the total deuterium error budget, as it is now dominated by the small error bars from the LUNA measurement\footnote{Some of these results were discussed in the 
\textit{Latest Advances in the Physics of BBN and Neutrino Decoupling Workshop} held virtually on 12-13 April 2021, see \href{https://indico.ph.tum.de/event/6798/}{https://indico.ph.tum.de/event/6798/}.}. However, these groups use different parametrisations and experimental measurements to fit the two other relevant reaction rates for the prediction of $\text{D/H}|_{\rm P}$, namely $d+d\to n +{}^{3}\text{He}$ and $d+d\to p +{}^{3}\text{H}$. In particular, Ref.~\cite{Pitrou:2020etk} fits these rates to theoretical models, Ref.~\cite{Pisanti:2020efz} uses a polynomial function to fit the rates, and Ref.~\cite{Yeh:2020mgl} uses the rates from the NACRE-II compilation~\cite{Xu:2013fha}. As a result of these differences in approach, the various groups report different predictions for $\text{D/H}|_{\rm P}$ at a fixed value of the baryon density $\Omega_\mathrm{b}h^2$. Using the best-fit value reported by the Planck collaboration for a $\Lambda$CDM cosmology, $\Omega_\mathrm{b}h^2 = 0.02236$~\cite{Planck:2018vyg}, the three groups obtain the following abundances:
\begin{align}\label{eq:DHvals}
\text{D/H}|_{\rm P} &=\left(2.49 \pm 0.11 \right)\times 10^{-5}\,,  \qquad \! [\text{Yeh et al. '21}] \\
\text{D/H}|_{\rm P} &= \left(2.52\pm 0.07 \right)\times 10^{-5}\,, \qquad  \! [\text{Pisanti et al. '21}] \\
\label{eq:DHPitrou}
\text{D/H}|_{\rm P} &= \left(2.45 \pm 0.04 \right)\times 10^{-5}\,. \qquad [\text{Pitrou et al. '21}] 
\end{align}
To assess the concordance between BBN and CMB determinations of the baryon density, these predictions should then be compared to the measured value of $\text{D/H}|_{\rm P}$, e.g. the one as recommended by the PDG~\cite{pdg}:  $\text{D/H}|_{\rm P}^{\rm obs} = (2.547\pm 0.025)\times 10^{-5}$. We can clearly appreciate that the results of Yeh et al.~\cite{Yeh:2020mgl} and Pisanti et al.~\cite{Pisanti:2020efz} are in agreement with this observed value. On the other hand, the results of Pitrou et al.~\cite{Pitrou:2020etk} show a slight tension between the predicted value of $\text{D/H}|_{\rm P}$ in Eq.~\eqref{eq:DHPitrou} and $\text{D/H}|_{\rm P}^{\rm obs}$, which can in turn be rephrased as a $1.6\sigma$ tension on the reconstructed value of $\Omega_\mathrm{b}h^2$ from BBN observations and the CMB~\cite{Pitrou:2020etk}. \\ 

\noindent \textbf{Updates in our BBN Analysis.} In Ref.~\cite{Sabti:2019mhn}, we used the BBN code \texttt{PRIMAT}~\cite{Pitrou:2018cgg,Pitrou:2019nub} linked to the cosmological code \texttt{NUDEC\_BSM}~\cite{Escudero:2018mvt,EscuderoAbenza:2020cmq} to calculate the evolution of the primordial element abundances in the presence of light thermal dark sectors. Ref.~\cite{Sabti:2019mhn} was submitted before the recent LUNA measurements were released, and as such, the calculations were carried out using the rates and uncertainties from Ref.~\cite{Pitrou:2018cgg}. Here, we comment on how our results are modified with the inclusion of the updated rates from each of the three groups. For this purpose, we again use \texttt{PRIMAT}, but we change the relevant nuclear reaction rates to those outlined in~\cite{Pitrou:2020etk}, \cite{Pisanti:2020efz}, and \cite{Yeh:2020mgl}. Then, at the level of our data analysis, we take the theoretical uncertainties in the predicted value of $\text{D/H}|_{\rm P}$ to be $4.4\%$, $2.8\%$ and $1.6\%$ respectively, see Eqs.~\eqref{eq:DHvals}$-$\eqref{eq:DHPitrou}. In addition, we also use the latest recommended value of the observed deuterium abundance from the PDG, which has been updated from $\text{D/H}|_{\rm P}^{\rm obs} = (2.569\pm 0.027)\times 10^{-5}$ to $\text{D/H}|_{\rm P}^{\rm obs} = (2.547\pm 0.025)\times 10^{-5}$. We note that this shift in the measured value is only at the level of $0.8\sigma$, and that the recommended value of the primordial helium abundance is the same in the 2018 and 2020 editions of the PDG review on BBN.\\

\begin{table*}[t]
\begin{center}
\resizebox{\textwidth}{!}{
{\def\arraystretch{1.35}
\begin{tabular}{l|lc|ccccc}
\hline\hline
\multirow{2}{*}{\textbf{Type}$\,$}	& \multicolumn{2}{c|}{\textbf{BSM Particle}}   	 &  \multicolumn{5}{c}{$\,\, $ \textbf{Current Constraints} $\,\,$}   \\
   &	 \multirow{1}{*}{Particle}    	 &   \multirow{1}{*}{g-Spin}  &   
   \multicolumn{1}{c}{$\,\,\,$ BBN $\,\,\,$}  &
   \multicolumn{1}{c}{$\,$ BBN+$\Omega_\mathrm{b} h^2$}  &
   \multicolumn{1}{c}{$\,\,\,$ Planck $\,\,\,$}  &
   \multicolumn{1}{c}{Planck+$H_0$ $\,$}  &   \multicolumn{1}{c}{$\,$ BBN+Planck$\,\,$}  \\ 
  \hline \hline
\parbox[t]{8mm}{\multirow{5}{*}{\rotatebox[origin=c]{90}{\textbf{Neutrinophilic}}}}   

& Majorana   & 2-F  & 2.2 & $ {\color{NavyBlue}{3.5}}\, \, \, {\color{BrickRed}{3.0}} \, \, \, {\color{ForestGreen}{3.0}} \, \, \, {\color{Plum}{2.9}}$  & 8.4 & 4.9 & $ \zero{\color{NavyBlue}{8.4}}\, \, \, \zero{\color{BrickRed}{8.4}} \, \,\, \zero{\color{ForestGreen}{7.1}}\, \, \, \zero{\color{Plum}{6.8}}$  \\ \cline{2-8} 

& Dirac  & 4-F  & 3.7 & $ {\color{NavyBlue}{6.4}}\, \, \, {\color{BrickRed}{5.6}} \, \, \, {\color{ForestGreen}{5.8}} \, \, \, {\color{Plum}{5.7}}$  & 11.3 & 8.0 & $ {\color{NavyBlue}{11.2}}\, \, \, {\color{BrickRed}{11.2}} \, \, \, {\color{ForestGreen}{10.0}} \, \, \, \zero{\color{Plum}{9.7}}$ \\ \cline{2-8} 

& Scalar   & 1-B  & 1.3 & $ {\color{NavyBlue}{1.7}}\, \, \, {\color{BrickRed}{1.5}} \, \, \, {\color{ForestGreen}{1.5}} \, \, \, {\color{Plum}{1.4}}$ & 5.6 & 1.6 & $ \zero{\color{NavyBlue}{5.6}}\, \, \, \zero{\color{BrickRed}{5.5}} \, \, \, \zero{\color{ForestGreen}{4.3}} \, \, \, \zero{\color{Plum}{4.0}}$ \\ 
\cline{2-8} 

& Complex Scalar  & 2-B  & 2.3 & $ {\color{NavyBlue}{3.7}}\, \, \, {\color{BrickRed}{3.2}} \, \, \, {\color{ForestGreen}{3.2}} \, \, \, {\color{Plum}{3.1}}$ & 8.5 & 5.1 & $ \zero{\color{NavyBlue}{8.5}}\, \, \, \zero{\color{BrickRed}{8.4}} \, \, \, \zero{\color{ForestGreen}{7.2}} \, \, \, \zero{\color{Plum}{6.9}}$ \\ \cline{2-8} 

& Vector  & 3-B  & 3.1 & $ {\color{NavyBlue}{5.3}}\, \, \, {\color{BrickRed}{4.6}} \, \, \, {\color{ForestGreen}{4.7}} \, \, \, {\color{Plum}{4.6}}$ & 10.1 & 6.8 & $ {\color{NavyBlue}{10.1}}\, \, \, {\color{BrickRed}{10.1}} \, \, \, \zero{\color{ForestGreen}{8.9}} \, \, \, \zero{\color{Plum}{8.6}}$ \\ 
\cline{2-8} 
\hline	

  \parbox[t]{8mm}{\multirow{5}{*}{\rotatebox[origin=c]{90}{\textbf{Electrophilic}}}}   

&Majorana & 2-F  &  0.5 & $ {\color{NavyBlue}{0.7}}\, \, \, {\color{BrickRed}{0.7}} \, \, \, {\color{ForestGreen}{2.9}} \, \, \, {\color{Plum}{3.3}}$ & 4.4 & 9.2 & $ \zero{\color{NavyBlue}{5.0}}\, \, \, \zero{\color{BrickRed}{4.7}} \, \, \, \zero{\color{ForestGreen}{7.1}}\, \, \, \zero{\color{Plum}{7.7}} $  \\ 
\cline{2-8} 

&Dirac   & 4-F  & 0.7 & $ {\color{NavyBlue}{4.2}}\, \, \, {\color{BrickRed}{3.5}} \, \, \, {\color{ForestGreen}{6.3}} \, \, \, {\color{Plum}{6.6}}$ & 7.4 & 12.0 & $ \zero{\color{NavyBlue}{8.0}}\, \, \, \zero{\color{BrickRed}{7.8}} \, \, \, {\color{ForestGreen}{10.0}} \, \, \, {\color{Plum}{10.5}}$   \\  
\cline{2-8} 

&Scalar  & 1-B  & 0.4 & $ {\color{NavyBlue}{0.4}}\, \, \, {\color{BrickRed}{0.4}} \, \, \, {\color{ForestGreen}{0.5}} \, \, \, {\color{Plum}{0.6}}$ & $\,\;2.4{}^\star$ & 6.4 & $ \zero{\color{NavyBlue}{1.6}}\, \, \, \zero{\color{BrickRed}{1.2}} \, \, \, \zero{\color{ForestGreen}{4.2}} \, \, \, \zero{\color{Plum}{4.8}}$   \\  \cline{2-8} 

&Complex Scalar   & 2-B  & 0.5 & $ {\color{NavyBlue}{0.9}}\, \, \, {\color{BrickRed}{0.8}} \, \, \, {\color{ForestGreen}{3.2}} \, \, \, {\color{Plum}{3.6}}$ & 4.6 & 9.2 & $ \zero{\color{NavyBlue}{5.1}}\, \, \, \zero{\color{BrickRed}{4.9}} \, \, \, \zero{\color{ForestGreen}{7.2}} \, \, \, \zero{\color{Plum}{7.8}}$  \\  \cline{2-8} 

&Vector    & 3-B  & 0.6 & $ {\color{NavyBlue}{3.0}}\, \, \, {\color{BrickRed}{2.3}} \, \, \, {\color{ForestGreen}{5.1}} \, \, \, {\color{Plum}{5.4}}$ & 6.3 & 10.9 &$ \zero{\color{NavyBlue}{6.9}}\, \, \, \zero{\color{BrickRed}{6.6}} \, \, \, \zero{\color{ForestGreen}{8.8}} \, \, \, \zero{\color{Plum}{9.4}}$   \\ 
\hline \hline
\end{tabular}
}
}
\end{center}

\caption{Lower bounds at 95.4\% CL on the masses of various thermal BSM particles in MeV. The columns correspond to analyses across different data sets, and the colors indicate our resulting constraints from taking the nuclear reaction rates used by {\color{NavyBlue}{Pisanti et al. '21}}~\cite{Pisanti:2020efz}, {\color{BrickRed}{Yeh et al. '21}}~\cite{Yeh:2020mgl}, {\color{ForestGreen}{Pitrou et al. '21}}~\cite{Pitrou:2020etk} or {\color{Plum}{Pitrou et al. '18}}~\cite{Pitrou:2018cgg}. The bounds for the BBN, Planck and Planck+$H_0$ analyses are insensitive to the choice of nuclear reaction rates between these groups. In Ref.~\cite{Sabti:2019mhn}, we used the rates from {\color{Plum}{Pitrou et al. '18}} and refer the reader to this reference for a detailed description of the data set used in each case. ${}^\star$This bound is only at 86\% CL.}\label{tab:DMbounds_new}
\end{table*}

\noindent \textbf{Updated Bounds on Neutrinophilic and Electrophilic Particles.} In Table~\ref{tab:DMbounds_new}, we show the resulting constraints on the masses of electrophilic and neutrinophilic particles coupled to the Standard Model bath. In particular, we show the resulting constraints that arise from using the three different groups' determinations of the nuclear reaction rates for $d+p\to {}^3\text{He} + \gamma$, $d+d\to n +{}^{3}\text{He}$, and $d+d\to p +{}^{3}\text{H}$. 

First of all, we note that these nuclear reaction rates do not alter the predicted value of the cosmological helium abundance $Y_{\rm P}$. This means that our Planck and Planck+$H_0$ analyses are unaltered by these updates. Secondly, we note that the primordial deuterium abundance is strongly sensitive to the baryon energy density, $\text{D/H}|_{\rm P} \propto (\Omega_\mathrm{b} h^2)^{-1.6}$, while the primordial helium abundance is largely insensitive to it, $Y_{\rm P} \propto (\Omega_\mathrm{b} h^2)^{0.04}$~\cite{Fields:2019pfx}. This means that in the BBN-only analysis, which solely includes data from $Y_{\rm P}|^{\rm obs}$ and $\text{D/H}|_{\rm P}^{\rm obs}$, the role of $\text{D/H}|_{\rm P}^{\rm obs}$ is to determine the value of $\Omega_\mathrm{b}h^2$, while the measured helium abundance is the driving power behind constraints on the masses of light dark sector states (see also Figure~5 in~\cite{Sabti:2019mhn}). As a result, our constraints for the BBN-only analysis are insensitive to the updated deuterium rates. This is the reason why there is only a single number in the corresponding columns for BBN, Planck, and Planck+$H_0$ in Table~\ref{tab:DMbounds_new}. Note that compared to Table II in~\cite{Sabti:2019mhn}, the BBN numbers in Table~\ref{tab:DMbounds_new} have changed by up to $0.1\,\text{MeV}$ for neutrinophilic particles as a result of the updated value we use for the observed deuterium abundance. 

In addition, we considered a separate analysis that included a very weak prior on the baryon energy density from CMB observations ($\Omega_\mathrm{b}h^2 = 0.02225\pm 0.00066$), together with measurements of $Y_{\rm P}$ and $\text{D/H}|_{\rm P}$. For this data set combination, we find that the $2\sigma$ constraints on neutrinophilic states do not depend appreciably on the nuclear reaction rates used. On the other hand, the bounds on electrophilic states are sensitive to the choice of nuclear reaction rates. In particular, we find that using the rates of Pisanti et al. '21~\cite{Pisanti:2020efz} or Yeh et al. '21~\cite{Yeh:2020mgl} leads to constraints that are up to $2-3$ MeV weaker than the ones that are obtained when using the rates from Pitrou et al. '21~\cite{Pitrou:2020etk} or '18~\cite{Pitrou:2018cgg}. The main reason for this difference in the constraints stems from the fact that the predicted value of $\text{D/H}|_{\rm P}$ by Pitrou et al., computed in the SM for the best-fit value of the baryon density from Planck (see Eq.~\eqref{eq:DHPitrou}), is ${\sim}2\sigma$ smaller than the observed value. Since the presence of electrophilic particles with masses $m_\chi \lesssim 10\,\text{MeV}$ decreases $\text{D/H}|_{\rm P}$
with respect to its SM value (see Figure 1 in~\cite{Sabti:2019mhn}), the predicted value of $\text{D/H}|_{\rm P}$ is even further in tension with the observed value when calculated at the baryon densities inferred from CMB observations. 
As a result of this tension, the bounds obtained by using the rates of Pitrou et al. '21 are stronger than those from Pisanti et al. '21 or Yeh et al. '21 in these cases. Note that we also consider an analysis that combines all BBN and Planck data. In this case, we again observe a similar variation in the constraints across the three groups, which has an identical explanation to that found in the BBN+$\Omega_\mathrm{b}h^2$ analysis.

Finally, we would like to comment on which bounds to use. Clearly, there is no issue for the constraints that are insensitive to the set of nuclear reaction rates used to derive them. However, for the BBN+$\Omega_\mathrm{b}h^2$ and BBN+Planck analyses, we believe that any of the constraints based on the rates from the three different groups~\cite{Pisanti:2020efz,Yeh:2020mgl,Pitrou:2020etk} represent a valid and meaningful bound, with any difference between them simply highlighting the current level of theoretical uncertainty in the prediction of the primordial deuterium abundance\footnote{At the moment, this theoretical uncertainty arises due to the lack of detailed knowledge of the $d+d\to n +{}^{3}\text{He}$ and $d+d\to p +{}^{3}\text{H}$  nuclear reaction rates. Therefore, it would be very useful to measure them more precisely, as that could lead to improved cosmological tests for several extensions of the Standard Model~\cite{Pitrou:2021vqr}.}. Of course, since these bounds are relevant for many scenarios beyond the Standard Model, one could take a conservative approach and use the weakest bound for any state. We would like to stress, however, that although the bounds in Table~\ref{tab:DMbounds_new} are derived at $2\sigma$, BBN and CMB observations disfavour dark sector states with even lighter masses at more than $5\sigma$, see e.g. Figure 2 in~\cite{Sabti:2019mhn}. We believe that it is important to emphasise that this conclusion is independent of the set of nuclear reaction rates used in the BBN analysis. \\

\noindent \textbf{Summary.} In conclusion, our bounds on MeV-scale dark sectors from our BBN-only and CMB-only analyses in~\cite{Sabti:2019mhn} are unchanged by the recent results from the LUNA collaboration. Similarly, we find that the bounds on neutrinophilic species are largely insensitive to the choice of nuclear reaction rates, even when BBN and CMB data are combined in the analysis. For electrophilic species, however, we do find that the $2\sigma$ bounds coming from combined BBN and CMB analyses can vary by up to $2-3\,\mathrm{MeV}$. The exact variation depends on the particular data set used, as well as the fermionic or bosonic nature of the dark sector state under consideration.\\

\renewcommand{\baselinestretch}{0.75}\normalsize
\renewcommand{\baselinestretch}{1.0}\normalsize

\noindent \textbf{Acknowledgments.} ME is supported by a Fellowship of the Alexander von Humboldt Foundation. NS is a recipient of a King's College London NMS Faculty Studentship.  In the early part of this work, ME and MF were supported
by the European Research Council under the European Union’s
Horizon 2020 program (ERC Grant Agreement No 648680 DARKHORIZONS).
In addition, the work of MF was supported partly by
the STFC Grant ST/P000258/1.  DB is supported by a `Ayuda Beatriz Galindo Senior' from the Spanish `Ministerio de Universidades', grant BG20/00228. IFAE is partially funded by the CERCA program of the Generalitat de Catalunya.

\newpage
\bibliographystyle{JHEP}
\bibliography{biblio}

\providecommand{\href}[2]{#2}\begingroup\raggedright\begin{thebibliography}{10}

\bibitem{Mossa:2020gjc}
V.~Mossa et~al., \emph{{The baryon density of the Universe from an improved
  rate of deuterium burning}},
  \href{https://doi.org/10.1038/s41586-020-2878-4}{\emph{Nature} {\bfseries
  587} (2020) 210}.

\bibitem{Fields:2019pfx}
B.~D. Fields, K.~A. Olive, T.-H. Yeh and C.~Young, \emph{{Big-Bang
  Nucleosynthesis after Planck}},
  \href{https://doi.org/10.1088/1475-7516/2020/03/010}{\emph{JCAP} {\bfseries
  03} (2020) 010} [\href{https://arxiv.org/abs/1912.01132}{{\ttfamily
  1912.01132}}].

\bibitem{Pitrou:2020etk}
C.~Pitrou, A.~Coc, J.-P. Uzan and E.~Vangioni, \emph{{A new tension in the
  cosmological model from primordial deuterium?}},
  \href{https://doi.org/10.1093/mnras/stab135}{\emph{Mon. Not. Roy. Astron.
  Soc.} {\bfseries 502} (2021) 2474}
  [\href{https://arxiv.org/abs/2011.11320}{{\ttfamily 2011.11320}}].

\bibitem{Pisanti:2020efz}
O.~Pisanti, G.~Mangano, G.~Miele and P.~Mazzella, \emph{{Primordial Deuterium
  after LUNA: concordances and error budget}},
  \href{https://doi.org/10.1088/1475-7516/2021/04/020}{\emph{JCAP} {\bfseries
  04} (2021) 020} [\href{https://arxiv.org/abs/2011.11537}{{\ttfamily
  2011.11537}}].

\bibitem{Yeh:2020mgl}
T.-H. Yeh, K.~A. Olive and B.~D. Fields, \emph{{The impact of new
  $d(p,\gamma)$3 rates on Big Bang Nucleosynthesis}},
  \href{https://doi.org/10.1088/1475-7516/2021/03/046}{\emph{JCAP} {\bfseries
  03} (2021) 046} [\href{https://arxiv.org/abs/2011.13874}{{\ttfamily
  2011.13874}}].

\bibitem{Xu:2013fha}
Y.~Xu, K.~Takahashi, S.~Goriely, M.~Arnould, M.~Ohta and H.~Utsunomiya,
  \emph{{NACRE II: an update of the NACRE compilation of
  charged-particle-induced thermonuclear reaction rates for nuclei with mass
  number $A < 16$}},
  \href{https://doi.org/10.1016/j.nuclphysa.2013.09.007}{\emph{Nucl. Phys. A}
  {\bfseries 918} (2013) 61} [\href{https://arxiv.org/abs/1310.7099}{{\ttfamily
  1310.7099}}].

\bibitem{Planck:2018vyg}
{\scshape Planck} collaboration, N.~Aghanim et~al., \emph{{Planck 2018 results.
  VI. Cosmological parameters}},
  \href{https://doi.org/10.1051/0004-6361/201833910}{\emph{Astron. Astrophys.}
  {\bfseries 641} (2020) A6}
  [\href{https://arxiv.org/abs/1807.06209}{{\ttfamily 1807.06209}}].

\bibitem{pdg}
{\scshape Particle Data Group} collaboration, P.~Zyla et~al., \emph{{Review of
  Particle Physics}}, \href{https://doi.org/10.1093/ptep/ptaa104}{\emph{PTEP}
  {\bfseries 2020} (2020) 083C01}.

\bibitem{Sabti:2019mhn}
N.~Sabti, J.~Alvey, M.~Escudero, M.~Fairbairn and D.~Blas, \emph{{Refined
  Bounds on MeV-scale Thermal Dark Sectors from BBN and the CMB}},
  \href{https://doi.org/10.1088/1475-7516/2020/01/004}{\emph{JCAP} {\bfseries
  01} (2020) 004} [\href{https://arxiv.org/abs/1910.01649}{{\ttfamily
  1910.01649}}].

\bibitem{Pitrou:2018cgg}
C.~Pitrou, A.~Coc, J.-P. Uzan and E.~Vangioni, \emph{{Precision big bang
  nucleosynthesis with improved Helium-4 predictions}},
  \href{https://doi.org/10.1016/j.physrep.2018.04.005}{\emph{Phys. Rept.}
  {\bfseries 754} (2018) 1} [\href{https://arxiv.org/abs/1801.08023}{{\ttfamily
  1801.08023}}].

\bibitem{Pitrou:2019nub}
C.~Pitrou, A.~Coc, J.-P. Uzan and E.~Vangioni, \emph{{Precision Big Bang
  Nucleosynthesis with the New Code PRIMAT}},
  \href{https://doi.org/10.7566/JPSCP.31.011034}{\emph{JPS Conf. Proc.}
  {\bfseries 31} (2020) 011034}
  [\href{https://arxiv.org/abs/1909.12046}{{\ttfamily 1909.12046}}].

\bibitem{Escudero:2018mvt}
M.~Escudero, \emph{{Neutrino decoupling beyond the Standard Model: CMB
  constraints on the Dark Matter mass with a fast and precise $N_{\rm eff}$
  evaluation}},
  \href{https://doi.org/10.1088/1475-7516/2019/02/007}{\emph{JCAP} {\bfseries
  02} (2019) 007} [\href{https://arxiv.org/abs/1812.05605}{{\ttfamily
  1812.05605}}].

\bibitem{EscuderoAbenza:2020cmq}
M.~Escudero~Abenza, \emph{{Precision early universe thermodynamics made simple:
  $N_{\rm eff}$ and neutrino decoupling in the Standard Model and beyond}},
  \href{https://doi.org/10.1088/1475-7516/2020/05/048}{\emph{JCAP} {\bfseries
  05} (2020) 048} [\href{https://arxiv.org/abs/2001.04466}{{\ttfamily
  2001.04466}}].

\bibitem{Pitrou:2021vqr}
C.~Pitrou, A.~Coc, J.-P. Uzan and E.~Vangioni, \emph{{Resolving conclusions
  about the early Universe requires accurate nuclear measurements}},
  \href{https://doi.org/10.1038/s42254-021-00294-6}{\emph{Nature Rev. Phys.}
  {\bfseries 3} (2021) 231} [\href{https://arxiv.org/abs/2104.11148}{{\ttfamily
  2104.11148}}].

\end{thebibliography}\endgroup

\end{document}